\documentstyle[11pt]{article}

\newcommand{\CMP}[1]{{\em Commun. Math. Phys.} {\bf {#1}}}
\newcommand{\JMP}[1]{{\em J.~Math. Phys.} {\bf {#1}}}
\newcommand{\NP}[1]{{\em Nucl.Phys.~B} {\bf {#1}}}

\newcommand{\PL}[1]{{\em Phys. Lett.} {\bf {#1}}}

\newcommand{\LMP}[1]{{\em Lett. Math. Phys.} {\bf {#1}}}

\newcommand{\Map}{C_0^\infty}

\newcommand{\SU}{{\rm SU}}
\newcommand{\su}{{\rm su}}

\newcommand{\eq}{\begin{equation}}
\newcommand{\eqend}{\end{equation}}
\newcommand{\eqa}{\begin{eqnarray}}
\newcommand{\nonueqa}{\begin{eqnarray*}}
\newcommand{\eqaend}{\end{eqnarray}}
\newcommand{\nonueqaend}{\end{eqnarray*}}

\newcommand{\bma}[1]{\begin{array}{#1}}
\newcommand{\ema}{\end{array}}
\newcommand{\bc}{\begin{center}}
\newcommand{\ec}{\end{center}}

\newcommand{\Ref}[1]{(\ref{#1})}

\newcommand{\ee}[1]{\,\mbox{{\rm e}}^{#1}}
\newcommand{\ii}{{\rm i}}

\renewcommand{\phi}{\varphi}

\newcommand{\eps}{\varepsilon}

\newcommand{\sign}{{\rm sign}}

\newfam\msbfam
\batchmode\font\twelvemsb=msbm10 scaled\magstep1 \errorstopmode
\ifx\twelvemsb\nullfont\def\Bbb{\bf}
\else	\catcode`\@=11
	\font\tenmsb=msbm10 \font\sevenmsb=msbm7 \font\fivemsb=msbm5
	\textfont\msbfam=\tenmsb
	\scriptfont\msbfam=\sevenmsb \scriptscriptfont\msbfam=\fivemsb
	\def\Bbb{\relax\ifmmode\expandafter\Bbb@\else
 		\expandafter\nonmatherr@\expandafter\Bbb\fi}
	\def\Bbb@#1{{\Bbb@@{#1}}}
	\def\Bbb@@#1{\fam\msbfam\relax#1}
	\catcode`\@=\active
\fi
\newcommand{\R}{{\Bbb R}}
\newcommand{\C}{{\Bbb C}}

\newcommand{\N}{{\Bbb N}}

\newcommand{\f}{\frac}
\newcommand{\cA}{{\cal A}}
\newcommand{\cL}{{\cal L}}
\newcommand{\cH}{{\cal H}}
\newcommand{\cF}{{\cal F}}

\newcommand{\ccr}[2]{{[} {#1},{#2} {]} }        

\newcommand{\Tra}[1]{{\rm Tr} \left({#1}\right)}          
\newcommand{\Trareg}{{\rm Tr}_{reg}}
\newcommand{\TraC}[1]{{\rm Tr}_C \left({#1}\right)}          
\newcommand{\tra}[1]{{\rm tr} \left({#1}\right)}          
\newcommand{\trac}[1]{{\rm tr}_{N} \left({#1}\right)}          

\newcommand{\dd}{{\rm d}}
\newcommand{\D}{D\!\!\!\!\slash}
\newcommand{\ddd}{\hat{\rm d}}
\newcommand{\CC}{{\cal C}}
\newcommand{\hCC}{\hat{\cal C}}

\newcommand{\gN}{{\rm gl}_N}
\newcommand{\homega}{\hat\omega}
\newcommand{\hint}{\hat{\int}}
\newcommand{\dGam}{\dd\Gamma}

\begin{document}
\begin{flushright}
\today
\end{flushright}

\vspace{.4cm}
\begin{center}
{\Large \bf On anomalies and noncommutative geometry}\\
\vspace{0.3 cm}
{\large Edwin Langmann}\\
\vspace{0.3 cm}
{\em Theoretical Physics, Royal Institute of Technology, S-10044 Sweden}
\end{center}

\subsection*{(1) Introduction}
In the following I discuss examples where basic structures from
noncommutative geometry (NCG) \cite{C} naturally arise in quantum field
theory.  The discussion is based on work with the ultimate aim to get
better mathematical understanding of quantum gauge theory models like
QCD(3+1).  (There is also a close connection with the representation theory
of infinite dimensional Lie groups which I shall not discuss.) The examples
are restricted to external field problems, i.e.\ fermions coupled to
non--dynamical Yang--Mills fields.  This simplification makes possible a
complete mathematical analysis.  Though rather drastic, it already allows
to study in detail several non--trivial aspects of QFT like the structure
of UV divergences in the fermion sector and how they lead to anomalies.
Moreover, it motivates the development of new efficient calculation tools
which, as I believe, should also be useful for analyzing the fully
quantized theories.

The philosophy of NCG --- to generalize the differential geometric
machinery to situations without underlying manifold but rather algebras of
Hilbert space operator --- seems to be the natural way to understand the
relation between the rich differential geometric structure of
anomalies (anomalies as de Rham forms, characteristic classes, descendent
equations relating anomalies in different dimensions etc.) and their
explicit QFT derivation (`dirty' calculations using Feynman diagrams,
perturbation theory etc.).  A general idea here is to interpret Feynman
diagrams as regularized traces of certain operators on some Hilbert space,
and to try to identify NCG structures based on the algebra of these
operators.  The regularized traces are of operators which are not trace
class.  In the examples discussed anomalies can be identified as
regularized traces of commutators $\ccr{a}{b}=ab-ba$ of certain operators
$a$ and $b$ (I believe this is true in general).
Even though such an expression is always zero if e.g.\ $a$ is
trace class and $b$ bounded, it can still be defined in more general cases
and be non--zero then.  Such regularized traces $\Trareg(\ccr{a}{b})$
are also closely related to
the Wodzicki residue and the Dixmier trace playing a fundamental role in
NCG \cite{C}.

\subsection*{(2) NCG and Schwinger Terms}

{\bf Graded Differential Algebra (GDA).}
A basic object in NCG is a GDA
generalizing the notion of de Rham forms.  To motivate this notion we
recall the following purely algebraic characterization of de Rham forms on
$\R^d$ (for simplicity we restrict ourselves to manifolds $\R^d$, and all
our mappings are $C_0^\infty$, i.e.\ smooth and compactly supported).  One
starts with the algebra $\CC_d^{(0)}\equiv \Map(\R^d,
\gN)$ of $N\times N$--matrix valued
functions on $\R^d$.  With $\dd$ the usual exterior differentiation, one
defines $\CC_d^{(n)}$ as the space of all $n$--forms which are linear
combinations of $\omega_n=X_0\dd X_1\cdots \dd X_n$ with
$X_i\in\CC_d^{(0)}$, and
$\CC_d=\bigoplus_{n=0}^{\infty} \CC_d^{(n)}$
($\CC_d^{(n)}=\emptyset$ for $n>d$ here).  Then $\dd$ defines a
mapping $\CC_d^{(n)}\to\CC_d^{(n+1)}$ with $\dd^2=0$, and
\eq
\label{GDA}
\omega_n \omega_m\in\CC_d^{(n+m)},\quad
\dd(\omega_n \omega_m)=\dd(\omega_n)\omega_m + (-)^n \omega_n\dd \omega_m
\eqend
$\forall \omega_n\in\CC_d^{(n)}$, $\omega_m\in\CC_d^{(m)}, n,m\in\N_0 .$
Moreover, there a linear map $\int$ --- integration of de Rham forms ---
\eq
\int\omega_n = \left\{ \bma{cc} \int_{\R^d}\trac{\omega_n} & \mbox{ for
$n=d$} \\ 0 & \mbox{ otherwise} \ema \right. ,
\eqend
(${\rm tr}_N$ is the usual trace of $N\times N$--matrices)
and Stokes' theorem holds,  $\int\dd \omega = 0$ for all
$\omega\in\CC_d$. Such triple $(\CC_d,\dd,\int)$ is called a GDA.

An important example for a GDA based on algebras of
Hilbert space operators is as follows. Consider
a separable Hilbert space $\cH$ which is decomposed in two orthogonal
subspaces, $\cH=\cH_+\oplus \cH_-$. The operator $\eps$ which is $\pm 1$
on $\cH_\pm$ is a grading operator, $\eps^*=\eps^{-1}=\eps$, and $\cH_\pm =
\f{1}{2}(1\pm\eps) \cH$ ($*$ is the Hilbert space adjoint).
Denoting as $B$ and $B_1$ the bounded and trace class operators on $\cH$,
respectively, and as $B_{2p}=\{a\in B| (a^*a)^{p}\in B_1 \}$
(these are the so--called Schatten classes) one defines the algebras
\eq
g_p\equiv\left\{ u\in B| \ccr{\eps}{u} \in B_{2p}
\right\}
\eqend
for $2p$ a positive integer. Then
\eq
\label{nform}
\homega_n = (i)^n u_0\ccr{\eps}{u_1}\cdots
\ccr{\eps}{u_n}\quad \equiv u_0 \ddd u_1\cdots \ddd u_n
\quad \forall u_i\in g_p, n=0,1,\ldots
\eqend
can be regarded as generalized differential forms.
Indeed, denoting as $\hCC_p^{(n)}$ the space of all linear
combinations of $n$--forms \Ref{nform} ($\hCC_p^{(0)}=g_p$),
\eq
\ddd\homega_n = \ii(\eps \homega_n -(-)^n\homega_n\eps)
\eqend
defines a mapping $\hCC_p^{(n)}\to \hCC_p^{(n+1)}$ such that $\ddd^2 = 0$.
$\ddd$ can therefore can be regarded as exterior differentiation.  One can
easily check that the relations \Ref{GDA}--hat hold.
An integration $\hint$ can be defined as
\eq
\hat{\int} \hat\omega_n = \left\{ \bma{cc}
\TraC{\Gamma^{2p}\hat\omega_n} &\mbox{ for
$n= 2p-1$}
\\ 0& \mbox{ otherwise}\ema\right.
\eqend
($\Gamma^{2p}=\Gamma\, / \, 1$ for $2p$ odd$\,/\,$even) where $\Gamma$ is a
grading operator on $\cH$ such that $\eps\Gamma = -\Gamma\eps$ and
\eq
\label{trac}
\TraC{a}\equiv \f{1}{2}\Tra{a+\eps a\eps}
\eqend
is a conditional Hilbert space trace.
Stokes' theorem holds here due to cyclicity of trace.

Then $(\hCC_p,\ddd,\hint)$ with $\hCC_p =\bigoplus_{n=0}^\infty \hCC_p^{(n)}$
is a GDA. As discussed below, it actually is a natural generalization of the de
Rham complex $(\CC_d,\dd,\int)$ if $2p=d+1$.

{\bf Quasi--free Second Quantization (QFSQ).}
The abstract mathematical framework
which has been found useful for studying external field problems of
fermions is very much in the spirit of NCG, even though historically it has
been developed independently.  (For a more detailed discussion see e.g.\
\cite{QFSQ}; this approach mainly uses methods from functional analysis
\cite{RS1}. For an alternative approach based on differential geometric
methods see \cite{PSM}. The generalization of the latter from $g_1$ to
$g_{p\geq 1}$ was first given in \cite{MRM}. My discussion of this
is based on \cite{L1}.)

In an external field problem, the starting point is a 1-particle description
of the fermions, and the aim is to `second quantize' i.e.\ to find the
corresponding QFT description.  One has a Hilbert space $\cH$ describing the
possible 1-particle states, a Hamiltonian $H$ and other observables which
are given by self--adjoint operators on $\cH$.  Note in the following
that, even though in most applications $\cH$ is a $L^2$--space over
some space manifold, no reference to this manifold or the explicit
form of the observables is made, and this makes this framework very general
and flexible.

The Hamiltonian $H$ naturally defines a splitting of $\cH$ in the subspaces
of positive and negative energy states, $\cH=\cH_+\otimes \cH_-$.  The
corresponding grading operator $\eps$ with $\eps \cH_\pm = \pm \cH_\pm$ can be
written as $\sign(H)$ (using the spectral theorem of self--adjoint
operators where $\sign(x)=1$ ($-1$) for $x\geq 0$ ($x<0$)).

Given these data one can construct the corresponding QFT model.  The
fermions field algebra is defined as $C^*$--algebra generated by the field
operators $\psi^*(f)$ linear in $f\in \cH$ and $\psi(f)=\psi^*(f)^*$, obeying
the
CAR $\left(\psi^*(f) + \psi(g)\right)^2 = (f,g)$ (inner product in $\cH$).
The physical representation of this algebra is then on the fermion Fock
space $\cF$ over $\cH$ and is uniquely determined by the `vacuum' $\Omega$ such
that $\psi(f_+)\Omega =\psi^*(f_-)\Omega = 0$ for all $f_\pm\in \cH_\pm$,
which corresponds to the Dirac sea.

The aim then is to `second quantize' observables and construct for
operators $u$ on $\cH$ the corresponding multiparticle observables i.e.\
operators $\dGam(u)$ on $\cF$ such that
$\ccr{\dGam(u)}{\psi^*(f)}=\psi^*(uf)$.  One then finds that it is not
always possible to construct such an {\em operator} $\dGam(u)$ but only if
$u$ is in $g_1$ introduced above. Thus the Schatten ideal conditions of
NCG naturally appear here.  The $\dGam(u)$ for $u\in g_1$ form an algebra
of operators on $\cF$, and one has relations
\eq
\label{ca}
\ccr{\dGam(u)}{\dGam(v)} = \dGam(\ccr{u}{v}) + \hat{c}_1(u,v)
\eqend
where
\eq
\label{hatc1}
\hat c_1(u,v) = \f{1}{2}\TraC{u\ccr{\eps}{v}}
\eqend
is a term arising from the regularization (normal ordering) required in this
construction. It is called Schwinger term in the physics
literature and is a non--trivial 2--cocycle of the Lie algebra $g_1$.

As discussed below, this general framework is sufficient only for QFT in
1+1 dimensions since there the observables of interest actually are in
$g_1$.  In higher dimensions the interesting observables $u$ are only in
$g_{p\geq 2}$.  Then $\dGam(u)$ cannot be defined as operator.  It is,
however, still possible to define it as {\em sesquilinear form}.  (Recall
that for a s.l.f.  $A$ only transition amplitudes $(f,A g)$ are defined for
$f,g$ in some dense set of the Hilbert space.) To obtain suitable
generalization of \Ref{ca} one has to also consider
splittings $\cH$ given by other grading operators $F$, namely those for
which $F-\eps$ is in $g_p$.  We denote this set as $Gr_p$.  This is because
the unitary operators $U=\exp(\ii u)\in g_p$ generated by self--adjoint
operators $u\in g_p$, act as transformations changing $\eps$ to $F=U^{-1}\eps
U$ which are in $Gr_p$, $F-\eps = U^{-1}\ccr{\eps}{U}$.  Infinitesimally
this action is described by the Lie derivative $\hat \cL_u f(F)\equiv \left.
-\ii df(\exp(-\ii tu) F\exp(\ii tu))\right/dt|_{t=0}$ on quantities
$f$ depending on $F$.  The construction of $\dGam(u;F)$ can then be done by an
additional `wave function renormalization'.  One obtains an algebra of
operators $G(u,F) = \hat\cL_u + \dGam(u;F)$ similar to \Ref{ca} with a
Schwinger term $\hat c_{2p-1}(u,v;F)$ depending on $F\in Gr_p$, e.g.  for
$p=2$ \cite{MRM}
\eq
\label{hatc3}
\hat c_3(u,v;F) =
-\f{1}{8}\TraC{(F-\eps)\ccr{\ccr{\eps}{u}}{\ccr{\eps}{v}} } .
\eqend
The case $p=1$ is special since one can choose $\dGam(u;F)$ and $\hat c_1$
independent of $F$ and therefore can forget about $Gr_1$.

{\bf Gauss law anomalies.}
We now describe how the abstract framework of quasi--free second
quantization above is used to derive the anomalous commutators of Gauss law
generators for chiral QCD, i.e.\ chiral fermions coupled to a Yang--Mills
field.  Our setting is YM theory on $\R^d$ with structure group
$\SU(N)$ represented by $N\times N$--matrices (for simplicity we do not
distinguish $\SU(N)$ from its representation i.e.\ we assume
$\SU(N)\subset\gN$).  The space dimension is $d=1,3,5\ldots$.  We denote
as $\cA_d$ the set of all YM field configuration, i.e.\ 1-forms
$A=\sum_{i=1}^d A_i \dd x^i$ with $A_i\in\Map(R^d,\su(N))$
($\su(N)\subset\gN$ the Lie algebra of $\SU(N)$).  The gauge group is
$\Map(\R^d,\SU(N))$ and acts on $\cA$ as $A\mapsto U\circ A \equiv
U^{-1}AU -\ii U^{-1}\dd U$.  Its Lie algebra is
$\Map(\R^d,\su(N))$ acting on functionals $f$ of $A$ by the Lie derivative,
$\cL_X f(A) \equiv -\ii df(\ee{\ii tX}\circ A)/dt|_{t=0}$.

We now consider chiral fermions coupled to external Yang--Mills fields.
Cohomological arguments show that for odd dimensions $d$ there are
2--cocycles $c_d(X,Y;A)$, e.g.\
\eq
\label{c1}
c_1(X,Y) = \f{1}{2\pi}\int_{\R^1} \trac{X\dd Y}
\eqend
for $d=1$ \cite{PSM}
and
\eq
\label{c3}
c_3(X,Y;A) =\f{\ii}{24\pi^2} \int_{\R^3} \trac{A\ccr{\dd X}{\dd Y}}
\eqend
for $d=3$ \cite{FM}. It has been suggested on cohomological grounds
that these 2--cocycles should arise as Schwinger terms commutators of
the Gauss' law generators of chiral QCD \cite{FM}.
This can be shown using the general formalism of QFSQ described above.
(For a different solution to this problem for $d=3$ see \cite{M1}.)

The starting point is the 1-particle description of chiral fermions.
The states at fixed time are described by the Hilbert space
$h=L^2(\R^d)\otimes \C^\nu_{spin}\otimes\C^N_{color}$ where
$\nu=2^{(d-1)/2}$ is the number of spin indices.
For a given YM configuration $A$, the 1-particle Hamiltonian is
$\D_A=\sum_{i=1}^d\gamma^i\left(
-\ii\partial_i + A_i \right)$ with $\gamma^i$ the usual $\gamma$--matrices
acting on $\C^\nu_{spin}$ and obeying
$\gamma^i\gamma^j+\gamma^j\gamma^i=2\delta^{ij}$.
This naturally defines a self--adjoint operator on $h$ for all $A\in\cA_d$,
and so do all $X\in\Map(\R^d;\gN)$ (containing $\Map(\R^d,\SU(N))$ and
$\Map(\R^d,\su(N))$; we recall that every $X\in\Map(\R^d,\gN)$
defines a bounded operator on $h$,
$(Xf)(x)=X(x)f(x)$ for all $f\in h$, which we denote by the same
symbol).

The essential property now is that for $h$ and $\eps=\sign(\D_0)$ above,
there are natural embeddings of $\Map(\R^d,\gN)$ in $g_p$ and of $\cA_d$ in
$Gr_p$,
\eq
\label{embed}
\mbox{$X\in \Map(\R^d,\gN)$, $A\in\cA_d$ $\Rightarrow$
$X\in g_p$, $F_A\equiv \sign(\D_A)\in Gr_p$  for $2p = d+1$}.
\eqend
Thus QFSQ gives by restriction the algebra of Gauss law generators of
chiral QCD with Schwinger terms $\hat c_d(X,Y;F_A)$, and the question of
whether the Schwinger terms \Ref{c1} and \Ref{c3} arise in chiral QCD reduces
to the question whether the 2--cocycles \Ref{c1} and \Ref{hatc1} (for $d=1$)
and
\Ref{c3} and \Ref{hatc3} (for $d=3$) are cohomologous. This is very
nontrivial since the abstract Schwinger terms $\hat c_d$ are given by
highly non--local expressions whereas the Schwinger terms $c_d$ are local
integrals of de Rham forms.  For $d=1$ it has been known to be true since
quite some time \cite{QFSQ}, and for $d=3$ it was shown by explicit
calculation recently \cite{LM1}.  However, given our discussion of graded
differential algebras above, this result becomes very plausible since
$\hat c_d$ is (up to a constant) just the noncommutative generalization of
$c_d$.  For $d=1$ this is obvious, and for $d=3$ it follows if we
regard $F-\eps$ as the noncommutative generalization of $A$.  The latter,
however, is very natural: If the YM field is a pure gauge, $A=-\ii
U^{-1}\dd U$ we have $F_A-\eps =U^{-1}\eps U -\eps = -\ii U^{-1}\ddd U$.
Equivalence of Schwinger terms thus is just a special case of a general
embedding theorem of the de Rham complex $(\hCC_p,\ddd,\hint)$ in the
complex $(\CC_d,\dd,\int)$, especially that the noncommutative integral
$\hint$ generalizes integration of de Rham forms $\int$,
\eq
\label{int1}
(\ii)^d \TraC{\Gamma X_0\ccr{\eps}{X_1}\cdots \ccr{\eps}{X_d}} =
c_d \int_{\R}\tra{X_0\dd X_1 \cdots \dd X_d}
\eqend
$\forall X_i \in\Map(\R^d;\gN)$,
with some constants $c_d$ and $\Gamma=1$ for $d$ odd and
$\Gamma=\gamma_{d+1}$ for $d$ even.  A simple proof of this (motivated by
the calculation in \cite{LM1}) was recently given in \cite{L2} (one gets $c_d=
(2\ii)^{[d/2]}2\pi^{d/2} / d(2\pi)^d \Gamma(d/2)$).  For $d=1$ this proves
that the Schwinger terms are in fact identical (as $c_1=1/\pi$), for $d=3$ it
proves identity for pure YM field configurations which are pure gauges
(as $c_3 = \ii/3\pi^2$).

\subsection*{(3) Efficient Anomaly Calculations}
In \cite{LM1} we found that the calculus of pseudodifferential
operators is an extremely powerful calculational tool in anomaly
calculations: it gives a simple way of calculating regularized traces
$\Trareg$ of commutators of operators (these methods were used earlier
in a similar context in \cite{M1}). The same tool was essential in
\cite{L2}. Recently we used this very tool for a very short QFT
derivation of the axial anomalies for all even dimensions \cite{LM2}.  The
strategy was as usual, to write the effective action for fermions in an
external YM field as $\Trareg$ of a Dirac operator $\D_A$ and calculate its
variation under axial gauge transformations.  Using only elementary rules
for manipulating Hilbert space operators, we could write the latter as a
sum of terms $\Trareg(\ccr{a}{b})$.  To my opinion, this very short
derivation makes mathematically precise traditional perturbative
calculations of the anomaly and is very much in the spirit of NCG.

\subsection*{(4) Final Remarks}
In my discussion in {\bf (2)} I tried to convince the reader that the NCG
viewpoint (i.e.\ generalizing from the differential algebra
$(\CC_d,\dd,\int)$ to $(\hCC_p,\ddd,\hint)$) is very
useful for general QFT calculations concentrating on the
essential QFT aspects, namely the nature of the UV divergences
characteristic for a specific dimension. It naturally leads to
generalizations of Schwinger terms to the noncommutative setting.
This and the anomaly calculation discussed in {\bf (3)} suggest that all
YM fermion anomalies should have noncommutative generalizations.
I recently found that this is indeed the case. In fact, there is a
noncommutative
generalization of the whole tower of descent equations \cite{Zumino}.
It involves a generalization of the noncommutative integration formula
\Ref{int1} to $(d-n)$--dimensional submanifolds of $\R^d$ such that Stokes'
theorem holds (details will appear elsewhere). To my opinion, this
noncommutative
version of the descent equations provides a nice explanation of how the
rich geometric structure of anomalies arises from QFT: it is
present already on the level of Hilbert space operators entering
the Feynman diagrams.

\bc{\bf Acknowledgments}\ec
I would like to thank J. Mickelsson for a pleasant collaboration and
H. Grosse  and S. Rajeev for helpful discussions.


\end{document}